\documentclass[12pt,aps,prd,floatfix,nofootinbib,a4paper,nosuperscriptaddress]{revtex4-2}
\pdfoutput=1

\usepackage{amsmath, amssymb, amsfonts, amsthm, latexsym, epsfig, mathrsfs, xcolor, bbm, slashed, braket, thmtools}

\usepackage[all]{xy}

\usepackage[inline]{enumitem}

\usepackage{setspace}
\usepackage[marginal, multiple]{footmisc}

\usepackage{svg}

\usepackage[T1]{fontenc}
\usepackage[utf8]{inputenc}
\usepackage{lmodern}

\usepackage[colorlinks, allcolors=blue!70!black, linktocpage]{hyperref}

\numberwithin{equation}{section}

\usepackage{cleveref}

\usepackage{microtype}
\usepackage{subcaption}

\usepackage{floatrow}
\let\OLDtableofcontents\tableofcontents
\renewcommand\tableofcontents[1]{%
    {\baselineskip 0.5ex %
	\OLDtableofcontents{#1}}%
}

\let\OLDthebibliography\thebibliography
\renewcommand\thebibliography[1]{%
	\setstretch{1.079} 
	\OLDthebibliography{#1}%
	\small %
	\setlength{\itemsep}{0.2\baselineskip} 
}

\let\OLDfootnote\footnote
\renewcommand\footnote[1]{%
	\setlength{\footnotesep}{0.75\baselineskip}%
	{\footnotesize \OLDfootnote{#1}}%
}

\setlist[enumerate]{noitemsep, label=(\arabic*), ref=(\arabic*)}

\renewcommand\thesection{\arabic{section}}
\renewcommand\thesubsection{\arabic{subsection}}

\makeatletter
\def\p@subsection{\thesection.}
\def\p@subsubsection{\thesection.\thesubsection.}
\makeatother 

\theoremstyle{plain}

\theoremstyle{definition}

\declaretheorem[style=remark,qed=$\scriptstyle{\blacksquare}$,numberwithin=section]{remark} 

\creflabelformat{equation}{#2#1#3}

\crefname{section}{sec.}{sec.}
\crefname{appendix}{Appendix}{Appendices}
\crefname{figure}{Fig.}{Figs.}
\crefname{table}{Table}{Tables}

\crefname{definition}{Def.}{Defs.}
\crefname{prop}{Prop.}{Props.}
\crefname{lemma}{Lemma}{Lemmas}
\crefname{corollary}{Cor.}{Cors.}
\crefname{thm}{Theorem}{Theorems}
\crefname{remark}{Remark}{Remarks}

\crefname{ass}{Assumptions}{Assumptions}
\crefname{property}{Properties}{Properties}

\newcommand{\be}{\begin{equation}\begin{aligned}}
\newcommand{\ee}{\end{aligned}\end{equation}}

\newcommand{\mc}{\mathcal}

\newcommand{\mf}{\mathfrak}

\newcommand{\eqsp}{\, ,\quad} 

\newcommand{\defn}{\mathrel{\mathop:}=} 

\newcommand{\op}[1]{\boldsymbol{#1}}

\let\oldint\int
\renewcommand{\int}{\oldint\limits}

\let\oldlim\lim
\renewcommand{\lim}{\oldlim\limits}

\newcommand{\Hilb}{\mathscr{H}}

\newcommand{\antiHilb}{%
\hspace{4pt} 
  \vbox{%
    \hrule height 0.5pt
    \kern0.25ex
    \hbox{%
      \kern-0.3em
      \ifmmode\Hilb\else\ensuremath{\Hilb}\fi
      \kern0em
    }
  }
}

\begin{document}

\setstretch{1.2}

\title{Holography in the linearized quantum gravity regime and modular crossed product}

\author{Avinandan Mondal}
\email{avinandan@alumni.iitm.ac.in}
\affiliation{Raman Research Institute, Sadashivanagar, Bengaluru 560080, India.}


\begin{abstract}
Within the semi-classical regime of AdS/CFT correspondence, we consider the limit where the bulk dynamical field is linearized metric perturbations satisfying linearized Einstein equations over background pure AdS spacetime. AdS/CFT correspondence gives us a holographic map, which is an isometric embedding map of the GNS Hilbert space of linearized gravity in the bulk (w.r.t. the AdS-invariant vacuum) to the GNS Hilbert space of CFT in the boundary (w.r.t. the Minkowski-invariant vacuum). We assume that the map takes AdS-vacuum in the bulk to CFT-vacuum in the boundary and that it allows AdS-Rindler wedge reconstruction. Then using this map, we show that for a given ball-shaped region in the boundary $A$, the relative entropy of a bulk state w.r.t. the AdS vacuum in the algebra of causal wedge associated to $A$ matches with the relative entropy of the dual CFT state w.r.t. the CFT vacuum in the algebra of CFT observables in $A$ in the code subspace, which is known as Jafferis-Lewkowycz-Maldacena-Suh (JLMS) condition. Furthermore, for localized semi-classical coherent excitations in the causal wedge associated to $A$ which corresponds to perturbed bulk geometry, we show rigorously using modular crossed product construction that the state-dependent part of entropy of the dual CFT state in the dressed Type-II algebra associated to $A$ satisfies vacuum subtracted Hubeney-Rangamani-Takayanagi (HRT) formula. 
\end{abstract}

\maketitle
\newpage 
\tableofcontents

\section{Introduction}\label{sec:intro}

AdS/CFT correspondence \cite{MaldacenaOG, WittenHolo} is a duality between a theory of quantum gravity (typically string theory or M-theory) in bulk anti de-Sitter (AdS) spacetime and a dual conformal field theory (CFT) lying at the conformal boundary of the conformally compactified AdS spacetime. The CFT lying at the boundary is typically a gauge theory with a gauge group labelled by parameter $N$ (e.g. if it is a conformal Yang-Mills theory, then the gauge group is $SU(N)$, e.g. in \cite{MaldacenaOG}). For the case of boundary gauge theory, the coupling parameter is described by t'Hooft coupling $\lambda$ which is defined for Yang-Mills theories as $\lambda = g_{YM}^2N$ where $g_{YM}$ is Yang-Mills gauge coupling. At large $N$ limit (keeping $\lambda$ fixed), the bulk string loops are suppressed and the bulk is described by tree-level strings. Then taking $\lambda >> 1$ (at large $N$) suppresses the stringy corrections in the bulk (as $\lambda >> 1$ corresponds to AdS length scale $L$ to be $L >> l_s$ where $l_s$ is string length) and the bulk is described by supergravity (see Chapter-5 of \cite{UserGuide} for a comprehensive review). Further, one can consistently truncate to the Einstein sector of supergravity and consider the metric to be the only dynamical field in the bulk (and consider other fields in supergravity as background fields) to obtain the classical Einstein limit. In this limit, the classical asymptotically AdS solutions of Einstein's equations are dual to certain CFT states in the boundary. Now consider one such classical solution, for our case, let it be pure AdS. Now we take the linearized quantum gravity limit, where one has quantized linearized metric perturbations (free gravitons) over this classical background spacetime. These graviton states are dual to certain states of the boundary dual CFT. This is precisely the limit where we shall be working in this paper.

In the classical limit of AdS/CFT correspondence, it has been proposed \cite{RT,HRT} that for CFTs with a gravitational dual, the von Neumann entropy of a CFT state dual to a classical bulk geometry in some boundary subregion $A$ (which we shall take to be a ball-shaped region) in the classical limit is equal to a quarter of the area of the extremal area codimension-2 surface $\Gamma$ in the bulk which is homologous to $A$ \footnote{Surfaces $\Gamma$ and $A$ are homologous mean that there exist a codimension 1 surface $C$ in the bulk such that $P \cup \Gamma$ serves as its boundary. In other words, the surface $\Gamma$ is ``anchored" at $\partial A$ (i.e. $\partial A = \partial \Gamma$ with $A - \Gamma = \partial C$)}:
\be 
    S(\psi, A) = \frac{\text{Ar}[\Gamma, g_{ab}]}{4G_N}, \label{eq:intro-RT}
\ee 
with $G_N$ as Newton's constant and $\text{Ar}[\Gamma, g_{ab}]$ denoting the area of the extremal surface $\Gamma$ computed w.r.t the bulk asymptotically AdS metric $g_{ab}$. This surface is called a Ryu-Takayanagi (RT) surface in static geometries, or Hubeney-Rangamani-Takayanagi (HRT) surface in dynamical spacetimes. We shall refer to extremal surfaces as HRT surfaces. According to the HRT proposal \cite{HRT}, corresponding to a boundary subregion $A$, the HRT surface $\Gamma$ is a non-expanding (i.e. $\vartheta_{\pm}|_{\Gamma} = 0$ where $\vartheta_+$ and $\vartheta_-$ are expansions of outgoing and ingoing null congruences from $\Gamma$) codimension-2 surface homologous to $A$ and in case multiple such surfaces exist, it is the one with minimum area.    

Now when one considers quantized perturbations over a fixed classical background AdS spacetime (which we shall take to be a pure AdS), then the entropy of the corresponding dual CFT state in $A$ is given by the generalized entropy of the quantum extremal surface (QES) $\Gamma$ \cite{QES} which is the surface that extremizes the generalized entropy which in the limit of linearized metric perturbations is given by \cite{FLM, QES}:
\be
    S_{\text{gen}} = \frac{\text{Ar}[\Gamma, g_{ab}]}{4G_N^{\text{bare}}} + S_{\text{QFT}} 
    \label{eq:QES}
\ee
where $S_{\text{QFT}}$ is the (ill-defined) divergent bulk QFT entropy and $G_N^{\text{bare}}$ is bare Newton's constant. Now, Susskind and Uglum \cite{Susskind} conjectured (and proved for an infinitely massive Schwarzschild black hole) that the generalized entropy of black holes can be written as:
\be
    S_{\text{gen}} = \frac{Area}{4G_N^{\text{bare}}} + S_{\text{QFT}} = \frac{Area}{4G_N} 
\ee
where $Area$ is the area of the event horizon and $G_N^{\text{ren}}$ is the renormalized Newton's constant. This was later shown more rigorously in the context of bifurcate Killing horizons spacetimes \cite{Kudler-Flam} and hence applies to our case as for pure AdS, the extremal surface is indeed a bifurcation surface for the AdS-Rindler Killing horizon (see \cref{sec:geometry}) and the bulk QFT is linearized graviton theory. Thus we can take the QES in \cref{eq:QES} to be the HRT surface in \cref{eq:intro-RT} if we replace $G_N$ by $G_N^{\text{ren}}$ in \cref{eq:intro-RT}. Hence, finally in the semi-classical limit of linearized metric perturbations, one has:
\be 
    S(\psi, A) = \frac{\text{Ar}[\Gamma, g_{ab}]}{4}, \label{eq:HRT-QES}
\ee 
where we have put renormalized Newton's constant $G_N^{\text{ren}} = 1$. 

An important point to note in \cref{eq:HRT-QES} is that the area $\text{Ar}[\Gamma, g_{ab}]$ is computed w.r.t. the physical AdS metric $g_{ab}$ (and not the conformally scaled \textit{unphysical} metric) and hence is infinite as the surface extends all the way to the boundary where the physical metric diverges. This is not too surprising as even the von Neumann entropy of a state in a quantum field theory (QFT) localized in a subregion diverges owing to the fact that QFT algebra restricted to a subregion is a Type-III von Neumann factor. However then it might seem like \cref{eq:HRT-QES} is not saying anything as both the L.H.S. and R.H.S. are infinite. The general way to deal with this problem in the literature is to put a radial cut-off and then to compute the extremal area surface. The HRT proposal \cite{HRT} also says that after putting a radial cut-off, in the case of multiple extremal area surfaces homologous to same boundary region, one should take the global minimal area surface. It was later shown by Sorce \cite{Sorce} that the difference in area between two extremal surfaces homologous to same boundary region is cut-off independent. Furthermore \cite{Sorce} also showed that if the perturbed spacetime with metric $g_{ab}'$ satisfies a weaker version of Fefferman-Graham falloff conditions, then for a given boundary subregion, the area of perturbed spacetime (with metric $g_{ab}'$)'s extremal surface $\Gamma'$ and the area of un-perturbed spacetime's extremal surface $\Gamma_0$ is finite:
\be
    \text{Ar}[\Gamma', g_{ab}'] - \text{Ar}[\Gamma_0, g_{ab}] < \infty, \label{eq:finite-diff}
\ee
and hence the \textit{vacuum subtracted} entropy is finite. Now in the context of AdS/CFT correspondence, since the CFT vacuum state corresponds to pure AdS bulk and some other dual state of the CFT corresponds to some perturbed asymptotically AdS spacetime, so the vacuum subtracted entropy of the dual CFT state is finite. 

In this paper, we consider a one-parameter family of asymptotically AdS spacetimes with metric $g_{ab}(\lambda)$, where $g_{ab}(\lambda = 0) =: g_{ab}^{(0)}$ is pure AdS and we shall be interested only in first order metric perturbations $\delta g_{ab} = \frac{dg_{ab}}{d\lambda}|_{\lambda = 0}$. We consider a connected ball-shaped region $A$ at some constant global AdS time section at the conformal boundary. The HRT surface corresponding to $A$ will be $\Gamma_0$ and $\Gamma_{\lambda}$ for vacuum $\omega_0$ (corresponding to $g_{ab}^{(0)}$ bulk) and some other state $\omega$ (corresponding to $g_{ab}(\lambda)$ bulk) of CFT respectively. So denoting $\Delta S(\omega, A) = S(\omega, A) - S(\omega_0, A)$, the HRT formula proposes that:
\be
    \Delta S(\omega, A) = \frac{1}{4}\big(\text{Ar}[\Gamma_{\lambda}, g_{ab}(\lambda)] - \text{Ar}[\Gamma_0, g_{ab}^{(0)}]\big), \label{eq:intro-main}
\ee
where the R.H.S. is the vacuum subtracted area and hence finite (see \cref{eq:finite-diff}). We also mention one important subtlety here. The HRT surface $\Gamma_{\lambda}$ exists as an extremal surface at all orders in $\lambda$ for asymptotically AdS spacetimes satisfying null curvature condition due to maximin construction by Wall \cite{WallMaximin}. However, for this work it is assumed that $\Gamma_{\lambda}$ is perturbatively away from $\Gamma_0$. More precisely, it means that $\Gamma_{\lambda}$ is a smooth one parameter family of extremal surfaces corresponding to $g_{ab}(\lambda)$. This is true for small enough values of $\lambda$ and we exclusively work in that regime. 

The main contribution of the paper is to show that to leading order in $\lambda$, the vacuum subtracted area appearing in the R.H.S. of \cref{eq:intro-main} is the state-dependent part of the well-defined Type-II von Neumann entropy of a \textit{crossed product} CFT state $\omega$ obtained as the dual state to a \textit{classical-quantum} coherent state of linearized quantum gravity coupled to an asymptotic gravitational charge in the bulk (with linearized metric perturbations supported in the AdS-Rindler wedge corresponding to the ball-shaped boundary region $P$ at some constant time). Once we show this, the L.H.S. of expression \cref{eq:intro-main}, which is the difference of two ill-defined (UV-divergent) entropies can be defined rigorously as the appropriate Type-II von Neumann entropy of the classical-quantum coherent state. This will consitute a proof of HRT formula in the semi-classical regime. In the course of this proof, we also shall show that the relative entropy of a bulk state (not necessarily coherent) w.r.t. the bulk vacuum matches with the relative entropy of the dual CFT state w.r.t. the CFT vacuum. This condition was first proposed by Jafferis, Lewkowycz, Maldacena, and Suh in \cite{JLMS} and is known as the JLMS condition.

HRT formula has already been proved in the literature using path integral and replica trick method \cite{replica}. Also HRT formula has been proved using relative entropy at the first variation level in the context of $AdS_3/CFT_2$ correspondence in \cite{Verch}, where it was shown that the relative entropy of a boundary coherent state w.r.t. the CFT vacuum in an interval in the boundary is equal to the first variation of geodesic length  where the dual geometry to the coherent CFT state was taken to be a Banados geometry. However, our proof, which is only valid in the linearized quantum gravity regime, is constructed using the well-defined von Neumann entropy for Type-II factors and doesn't rely on putting a UV cut-off to evaluate continuum ill-defined quantities and then showing cut-off independence. Also our proof works for $AdS_{d+1}/CFT_d$ correspondence for any $d \geq 2$.

This work is motivated by a previous work of the author with Prabhu \cite{MP} in the context of asymptotically flat dynamical black holes where a semi-classical entropy formula for dynamical black holes was evaluated using modular crossed product and was related to the classical dynamical black hole entropy formula of \cite{HWZ}. The applicability of modular crossed product construction in the presence of a geometric modular flow motivated us to use it for linearized quantum gravity on AdS spacetimes in the context of AdS/CFT correspondence. 

The rest of the paper is organized as follows. In \cref{sec:geometry} we discuss the AdS-Rindler geometry, set up the notations that shall be used in the rest of the paper and also discuss the gauge conditions on metric perturbations that we shall impose. In \cref{sec:quantization} we briefly review quantization of linearized gravity on AdS-Rindler horizon and discuss the observables of interest, in particular the flux operator. In \cref{sec:KMS-proof} we shall prove that AdS invariant vacuum is KMS in AdS-Rindler wedge. We shall use that fact in \cref{sec:crossed-product-in-bulk} to establish that modular flow in the AdS-Rindler wedge is geometric. In \cref{sec:crossed-product-in-bulk} we shall also show that  taking perturbative gravitational constraints into account reduces the algebra in the wedge from Type-III to Type-II as the appropriate invariant factor turns out to be the crossed product algebra of the AdS-Rindler wedge with the modular automorphism group of vacuum. In \cref{sec:crossed-product-boundary} we introduce the holographic map and list the assumptions on the map that we would require for it to satisfy. Using the map, we induce an isomorphism between the bulk algebra and boundary algebra in the code subspace. We also induce a crossed product in the boundary via the holographic map. Also, in \cref{sec:JLMS} we prove the JLMS condition. Finally, in \cref{sec:HRT-proof} we consider states in the crossed-product algebra of the bulk obtained from coherent perturbations in the bulk coupled to a slowly-varying wave function in the auxiliary Hilbert space of the boundary charge (which was needed to do the crossed product). Then, using this state, we prove the HRT formula at the leading order in perturbation. Finally we outline some limitations and avenues for future exploration in \cref{sec:outro}. 

\textit{Notations and conventions:} We shall use ``mostly plus'' convention for spacetime metric signature. The tensor indices are all abstract indices following Wald \cite{WaldBook}. Also, lowercase latin indices ($a, b, ...$) would denote bulk indices, whereas uppercase latin indices ($A,B,...$) would denote tensor indices on the $\mathbb S^{d-1}$ cuts of the AdS-Rindler horizon. In the operator algebraic side, if $\Hilb$ denotes a Hilbert space, then $\mathcal L(\Hilb)$ denotes the set of linear operators on $\Hilb$ and $\mathcal B(\Hilb)$ denotes the set of bounded linear operators on $\Hilb$ (w.r.t. the operator norm induced from the Hilbert space norm). Also when we say $\sigma$ is a flow on some space $X$, we mean $\sigma: X \times \mathbb R \rightarrow X$ is the map and $\sigma_t : X \rightarrow X$ is the automorphism on $X$ for $t \in \mathbb R$.

\section{Anti de-Sitter-Rindler wedge, metric perturbations and gauge conditions} \label{sec:geometry}

In this work we shall be interested in a one parameter family of $(d+1)$-dimensional asymptotically AdS spacetimes with manifold $M$ and have metrics $g_{ab}(\lambda)$. The metric $g_{ab}^{(0)} := g_{ab}(\lambda = 0)$ is the pure $AdS_{d+1}$ spacetime which is defined as the universal cover of a connected component of the section:
\be
    -T_1^2 - T_2^2 + \sum_{i=1}^{d}X_i^2 = - L^2 
\ee
of $\mathbb{R}^{d,2}$ where $(T_1, T_2, X_1, \cdots , X_d)$ is a Cartesian chart on $\mathbb{R}^{d,2}$. Without the loss of generality, we shall take $L = 1$ (i.e. all lengths are measured in units of $L$). 
The conformally completed manifold has a conformal boundary $\partial M$ which is timelike and has the topology of $\mathbb S^{d-1} \times R$ and has a conformally flat Minkowski metric in $d$ dimensions. The one parameter family of manifolds with  metrics $g_{ab}(\lambda)$ are asymptotically AdS, implying that the conformal completions of these spacetimes have isomorphic conformal boundaries. 

\begin{figure}[htbp]
        \centering
        \includegraphics[width=0.5\linewidth]{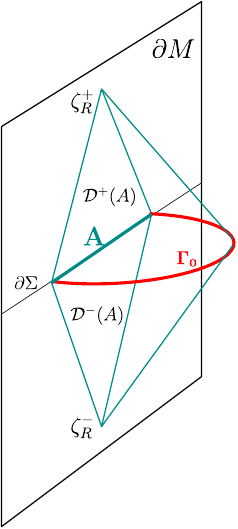}
        \caption{AdS viewed from the boundary: $\mc D^+(A)$ and $\mc D^-(A)$ are respective future and past domains of dependence of $A$ in the boundary $\partial M$. $\Gamma_0$ is the extremal surface anchored to $A$.}
        \label{fig:ads-boundary}
\end{figure}

\begin{figure}[htbp]
    \centering
    \includegraphics[width=0.7\linewidth]{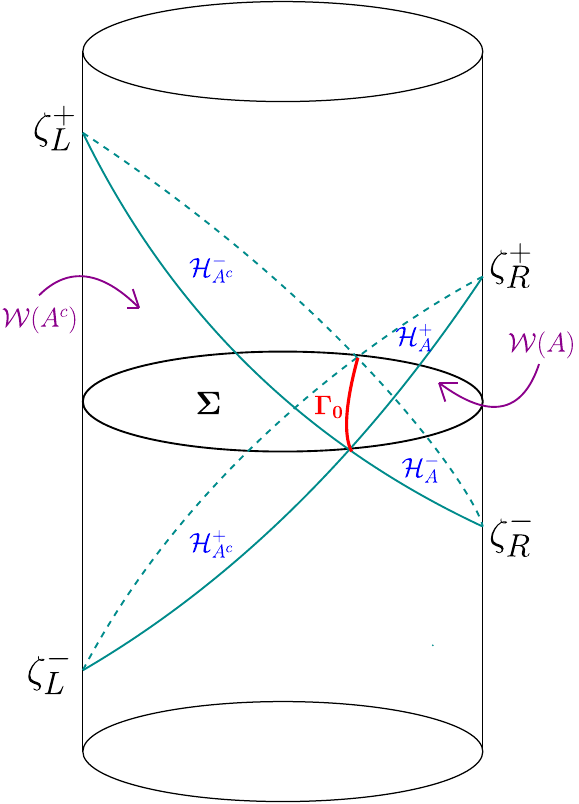}
    \caption{AdS spacetime: $A$ is the boundary subregion, with its causal wedge $\mathcal W(A)$. The causal wedge of its complement region $A^c = \partial\Sigma-A$ is $\mc W(A^c)$. The AdS-Rindler horizons $\mc H^+ \cup \mc H^-$ form a bifurcate Killing horizon structure w.r.t. Killing boosts $\xi^a$ where $\mc H^+ = \mc H^+_A \cup \mc H^+_{A^c}$ and $\mc H^- = \mc H^-_{A} \cup \mc H^-_{A^c}$. $\Gamma_0$ is the bifurcation surface and being non-expanding, it is the extremal surface anchored to $A$ (and $A^c$).}
    \label{fig:ads-main}
\end{figure}

Now consider the global coordinates of pure $AdS_{d+1}$ where the metric is written as:
\be
    ds^2 = -(1 + {r^2})dt^2 + \frac{dr^2}{(1 + {r^2})} + r^2d\Omega_{d-1}^2
\ee
and take a constant time slice $\Sigma$ ($t=0$ slice without the loss of generality). In the conformally completed spacetime, this space-like surface $\Sigma$ intersects the boundary $\partial M$ in a $\mathbb S^{d-1}$ section which is the boundary of $\Sigma$, i.e. $\partial\Sigma = \Sigma \cap \partial M$. Now consider a connected ball-shaped region $A \subset \partial \Sigma$ and its domain of dependence in the boundary which is the region $\mathcal D(A) = \mathcal{D}^+(A) \cup \mathcal D^-(A)$ as shown in the \cref{fig:ads-boundary}. Now consider the causal wedge $\mathcal W(A)$ defined as (see \cref{fig:ads-main}, \cref{fig:ads-boundary}):
\be
    \mc W(A) = J^-(\mc D(A)) \cap J^+(\mc D(A)).
\ee
The boundary of the causal wedge in the bulk is a union of two null surfaces $\mc H_{A}^+ \cup \mc H_{A}^-$, where $\mc H_{A}^-$ and $\mc H_{A}^+$ are \textit{past} and \textit{future} horizons associated to $A$, which are Killing horizons for AdS-Rindler boost field and hence the wedge $\mc W(A)$ shall also be referred to as the AdS-Rindler wedge associated to $A$. This AdS-Rindler boost field, henceforth denoted by $\xi^a$ is defined in the following paragraphs. 

The standard AdS-Rindler wedge is the causal wedge $\mc W(A_0)$ where $A_0$ is half of $\partial \Sigma$. This is isometric to the wedge $\mc W(A)$ with $A$ being any ball-shaped region by virtue of maximal symmetric nature of $AdS_{d+1}$. The standard AdS-Rindler boost vector can be written most explicitly by going to AdS-Rindler coordinates $(\tau, \sigma, Y^1,..., Y^{d-1})$ which is a coordinate patch in $\mc W(A_0)$ where $\tau$ is Rindler time, $\sigma$ acts as a radial coordinate and $(Y^1, ..., Y^{d-1})$ are hyperbolic coordinates on a $d-1$ dimensional hyperboloid. In these coordinates, the metric in $\mc W(A_0)$ can be written as:
\be
    ds^2 = -(\sigma^2 - 1)d\tau^2 + \frac{d\sigma^2}{\sigma^2 - 1} + \sigma^2dH_{d-1}^2
\ee
where $dH_{d-1}^2$ is the unit hyperboloid metric. Now from the above metric, it is clear that the boost field $\partial_{\tau}$ is Killing and it is timelike in $\mc W(A_0)$. 

\begin{figure}[htbp]
    \centering
    \includegraphics[width=0.6\linewidth]{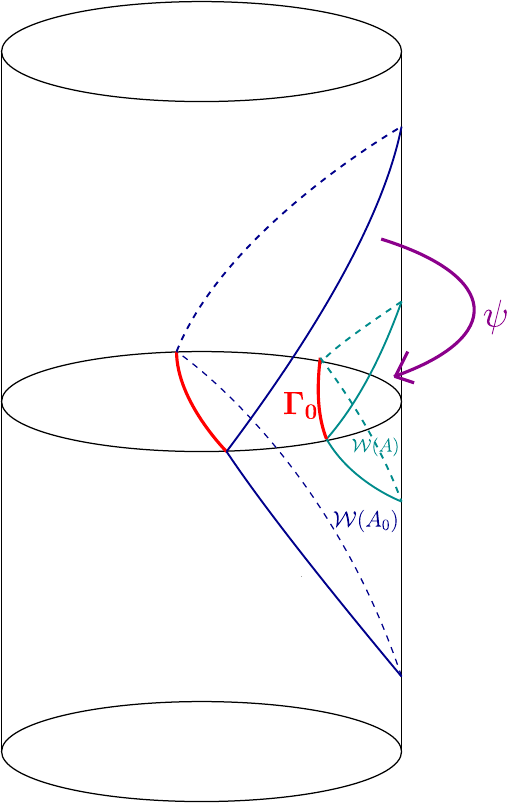}
    \caption{$\psi: AdS_{d+1} \to AdS_{d+1}$ is the isometry which maps $\mc W(A_0)$ to $\mc W(A)$.}
    \label{fig:ads-cft-isom}
\end{figure}

Let $\psi: AdS_{d+1} \to AdS_{d+1}$ is an isometry which maps $\mc W(A_0)$ to $\mc W(A)$ as shown in \cref{fig:ads-cft-isom}. The push-forward of the vector $\partial_{\tau}$ is the Killing vector $\xi^a$: 
\be
    \xi^a = \psi^*(\partial_{\tau}),
\ee
which is timelike in $\mc W(A)$ and generates the null surfaces $\mc H_{A}^{\pm}$. The null surfaces can be geodesically completed to $\mc H^{\pm}$ to generate a bifurcate Killing horizon structure w.r.t. the Killing field $\xi^a$ (where $\xi^a$ is extended to $\mc H^{\pm} - \mc H^{\pm}_A $ by taking $\xi^a$ to be the associated boost field in $\mc W(A^c)$) The surface $\Gamma_0 = \mc H^+_A \cap \mc H_A^-$ is the bifurcation surface (i.e. where $\xi^a = 0$). We shall denote the affine parameter on $\mc H^+$ by $V$ and Killing parameters on $\mc H^{+}_A$ and $\mc H^+_{A^c}$ by $v$ and $v'$ respectively. The surface gravity on the horizon $\mc H^+$ of $\xi^a$ can be evaluated to be $\kappa = 1$ and hence the affine and Killing parameters are related as:
\be
    V = e^{v} \quad (\text{on } \mc H^+_A), \\
    V = -e^{v'} \quad (\text{on } \mc H^+_{A^c}).
\ee

Now, assuming reflecting boundary conditions on $\partial M$, either of the horizons $\mc H^{\pm}$ can be used as an initial data surface for Cauchy evolution of field theories in the bulk. The codimension-2 bifurcation surface $\Gamma_0$ is the HRT surface corresponding to $A$. This is because an extremal surface anchored to some boundary ball-shaped region $A$ is a codimension 2 surface homologous to it and having both null expansions to be zero \cite{HRT} which is clearly the case for $\Gamma_0$. 

In this work we shall consider metric perturbations over this background pure AdS and consider the support of $\delta g_{ab}:= \frac{d}{d\lambda}g_{ab}|_{\lambda = 0}$ in the region $\mc W(A)$ and anywhere else where it propagates through the reflecting boundary (although higher order metric perturbations can be supported anywhere). In the perturbed spacetime $g_{ab}(\lambda)$, the HRT surface $\Gamma_{\lambda}$ corresponding to (the same boundary region) $A$ \footnote{Recall that $g_{ab}(\lambda)$ is a family of asymptotically AdS spacetimes and hence conformal boundary is unchanged} will be perturbed from $\Gamma_0$. But, we can choose Gaussian null gauge so that the perturbed spacetime's HRT surface $\Gamma_{\lambda}$ is $\mc O(\lambda^3)$ displaced from $\Gamma_0$. This has been elaborated in \cref{sec:gauge}.

\subsection{Gauge conditions on metric perturbations and their implications} \label{sec:gauge}

We impose Gaussian null gauge on metric perturbations \cite{BHBB} on the horizon $\mc H^+$. This ensures that the hypersurface $\mc H^+$ remain null at all orders in $\lambda$. Also we impose that the first order perturbed expansions vanish on $\Gamma_0$:
\be
    \delta\vartheta_{\pm}|_{\Gamma_0} = 0.
    \label{eq:gauge-cond}
\ee
which can be consistently imposed by exploiting residual gauge freedom after fixing Gaussian null gauge \cite{BHBB}. Hence, the expansion of $\Gamma_0$ in the metric $g_{ab}(\lambda)$ is of $\mc O(\lambda^2)$. So, the surface $\Gamma_0$ remains extremal at $\mc O(\lambda)$. Now say $\Gamma_{\lambda}$ is the true HRT surface which is assumed to be perturbatively away from $\Gamma_0$ (i.e. $\lambda$ defines a smooth one parameter family of extremal surfaces for $g_{ab}(\lambda)$). Now, say $\eta^a$ is the normal displacement vector field of the one-parameter family of surfaces $\Gamma_{\lambda}$ at $\lambda = 0$. Now denote the mean curvature vector of the embedded submanifold $\Gamma_0$ by $H^a$. Recall that the mean curvature vector $H^a$ of $\Gamma_0$ at some point $p \in \Gamma_0$ is defined as:
\be
    H^a(p) = \frac{1}{d-1}\Sigma_{i=1}^{d-1}\Pi^a(e_i, e_i)
\ee
where $\Pi^a$ is the second fundamental form tensor which takes two tangent vectors to $M$ and gives a vector normal to $\Gamma_0$. Also $(e_1, ..., e_{d-1})$ is any frame on $\Gamma_0$ at $p$. For more details on mean curvature vector refer to \cite{Barrett}. Now the change in area can be written as:
\be
    \text{Ar}[\Gamma_{\lambda}, g_{ab}(\lambda)] - \text{Ar}[\Gamma_0, g_{ab}(\lambda)] = \lambda (d-1)\int_{\Gamma_0} \epsilon^{d-1} H_a\eta^a + \lambda\int_{\Gamma_0} \epsilon^{d-1}\text{div}(\eta^T) + (\text{higher orders}),
    \label{eq:gauge-1}
\ee
where $\epsilon^{d-1}$ is the induced volume form on $\Gamma_0$ from $g_{ab}(\lambda)$. $\eta^T$ is the component of $\eta^a$ tangent to $\Gamma_0$ and $\text{div}(\eta^T)$ is its divergence in $\Gamma_0$. Now, the mean curvature vector $H^a$ can be written in terms of ingoing and outgoing null generators on $\Gamma_0$, namely $l^a$ and $n^a$ respectively and the respective expansions of the generated null surfaces at $\Gamma_0$ namely $\vartheta_-|_{\Gamma_0}$ and $\vartheta_+|_{\Gamma_0}$ as:
\be
    H^a = \frac{1}{d-1}\Big(l^a\vartheta_-|_{\Gamma_0} + n^a\vartheta_+|_{\Gamma_0}\Big) 
\ee
Due to the gauge condition \cref{eq:gauge-cond}, we have $H^a \sim \mc O(\lambda^2)$ and hence the first term in R.H.S. of \cref{eq:gauge-1} is $\mc O(\lambda^3)$. Now, by using Stokes' theorem on the second term in R.H.S. of \cref{eq:gauge-1} and using the fact that $\eta^T$ must vanish at $\partial\Gamma_0 = \partial A$ as $\partial\Gamma_{\lambda} = \partial\Gamma_0 = \partial A$ \footnote{A non-vanishing $\eta^T$ at $\partial \Gamma_0$ will lead to the surface $\Gamma_{\lambda}$ to detach from or pierce into the boundary which are not allowed as it is anchored to the boundary} we can say that (with $m^a$ being normal to $\partial\Gamma_0$ but tangent to $\Gamma_0$):
\be
    \int_{\Gamma_0}\epsilon^{d-1}\text{div}(\eta^T) = \int_{\partial\Gamma_0}\epsilon^{d-2}\eta^T_am^a = 0.
\ee
Thus, we conclude that:
\be
    \text{Ar}[\Gamma_{\lambda}, g_{ab}(\lambda)] - \text{Ar}[\Gamma_0, g_{ab}(\lambda)] \sim \mc O(\lambda^3).
    \label{eq:area-var-imp}
\ee
Now as $\mc H^+$ remains a null congruence at all orders in $\lambda$, so we can use Raychaudhuri's equation and at first order it reads:
\be
    \frac{d}{dV}(\delta\vartheta) = -\frac{2}{d-1}\vartheta^{(0)}\delta\vartheta - 2\sigma_{ab}^{(0)}\delta\sigma^{ab} - \delta R_{ab}n^an^b = 0,
    \label{eq:Raychaudhuri}
\ee
where the expression evaluates to zero since the background expansion and shear vanish $\vartheta^{(0)} = 0 = \sigma^{(0)}_{ab}$ (as $\mc H^+$ is a Killing horizons). Now the perturbations being source-free, by by virtue of source-free Einstein equations:
\be
    \delta R_{ab} = \big(-\Lambda + \frac{1}{2}R^{(0)}\big)\delta g_{ab} + \frac{1}{2}\delta R g_{ab}^{(0)}
\ee
where $\Lambda = - \frac{d(d-1)}{2}$ is the cosmological constant. Now Gaussian null gauge condition on $\mathcal H^+$ ensures that $g_{ab}(\lambda)n^an^b = 0$ at all orders in $\lambda$ and hence $\delta R_{ab}n^an^b = 0$. So, from \cref{eq:Raychaudhuri} and using the gauge condition \cref{eq:gauge-cond} we can say that the first order perturbed expansion vanishes on the entire horizon:
\be
    \delta\vartheta|_{\mc H^+} = 0.
    \label{eq:exp-vanish}
\ee
Hence, we have area of cuts of the horizon $\mc H^+$ to vary only at second order. So we have $\text{Ar}[\Gamma_0, g_{ab}(\lambda)] - \text{Ar}[\Gamma_0, g_{ab}^{(0)}]  \sim \mc O(\lambda^2)$ and we shall denote the second order piece of it by $\delta^2\text{Ar}[\Gamma_0]$.

Now, the change in area of the HRT surface of the perturbed spacetime compared to the HRT surface of the un-perturbed spacetime is:
\be
    \text{Ar}[\Gamma_{\lambda}, g_{ab}(\lambda)] - \text{Ar}[\Gamma_0, g_{ab}^{(0)}] \\ 
    =\text{Ar}[\Gamma_{\lambda}, g_{ab}(\lambda)] - \text{Ar}[\Gamma_0, g_{ab}(\lambda)] + \text{Ar}[\Gamma_0, g_{ab}(\lambda)] - \text{Ar}[\Gamma_0, g_{ab}^{(0)}] \\
    = \delta^2\text{Ar}[\Gamma_0] + \mc O(\lambda^3),
    \label{eq:principal-eq-area}
\ee
where we added and subtracted $\text{Ar}[\Gamma_0, g_{ab}(\lambda)]$ to go from first to second line and then subsequently used \cref{eq:area-var-imp} and the fact that $\mc O(\lambda^2)$ term in $\text{Ar}[\Gamma_0, g_{ab}(\lambda)] - \text{Ar}[\Gamma_0, g_{ab}^{(0)}]$ is $\delta^2\text{Ar}[\Gamma_0]$ to move to the third line.

\section{Algebraic quantization on AdS-Rindler horizon} \label{sec:quantization}

The horizon $\mc H^{+}$ behaves as an initial data surface for first-order (linearized) metric perturbations in $AdS_{d+1}$ (with reflecting boundary conditions at $\partial M$). The classical phase space construction for linearized gravity on $\mc H^+$ proceeds in a similar way as outlined in \cite{MP} and we refer the reader to \cite{MP} and references therein for review.

The phase space observable of central importance is the flux operator $F_{\xi}$ which is the Hamiltonian for the Killing time translation on the horizon $\mc H^+$:
\be
    F_{\xi} = \int_{\mc H^+}\omega(g; \delta g, \pounds_{\xi}g) = \frac{1}{4\pi}\int_{\mc H^+} dV d\Omega_2 \delta\sigma_{AB}\delta\sigma^{AB},
    \label{eq:flux-op-classical-1}
\ee
which can be written as a boundary integral:
\be
    F_{\xi} = \int_{\mc H^+}d(\delta^2Q_{\xi} - \xi\cdot\delta\theta(g, \delta g)),
    \label{eq:flux-op-classical-2}
\ee
where $\omega$ is the symplectic $d$-form, $Q_{\xi}$ is the Noether charge 2-form associated with Killing flow and $\theta$ is the symplectic potential $d$-form. Now denoting the intersection of $\mc H^+$ with $\partial M$ to be $\zeta_R^+$ and $\zeta_L^-$ (see \cref{fig:ads-main}), we can integrate \cref{eq:flux-op-classical-2} by parts to get:
\be
    F_{\xi} = X - C
\ee
where:
\be
    X := \int_{\zeta_R^+}(\delta^2Q_{\xi} - \xi\cdot\delta\theta(g, \delta g)) \\
    C := \int_{\zeta_L^-}(\delta^2Q_{\xi} - \xi\cdot\delta\theta(g, \delta g))
    \label{eq:boundary-charges-classical}
\ee
An important point to note is that although $F_{\xi}$ is a phase space observable of linearized gravity, but the boundary charges $X$ and $C$ individually are not phase space observables since they have a dependence on second order metric perturbations $\delta^2 g$.

On $\mc H^{+}$ we quantize the smeared perturbed shear $\delta\sigma(s)$ as outlined in \cite{MP, Gautam1, Gautam2}. We do GNS construction of the algebra $\mc A_{\mc H^+}$ generated by smeared shear operators $\op{\delta\sigma}(s)$ on $\mc H^{+}$ w.r.t. the global AdS isometry invariant vacuum $\omega_0$ and obtain GNS Hilbert space $\Hilb_{\text{AdS}}$, a cyclic separating state representing the AdS vacuum $|\omega_0\rangle$ and a representation of the algebra $\mc A_{\mc H^+}$ in $\mc L(\Hilb_{\text{AdS}})$. The flux operator is now a QFT observable:
\be
    \op F_{\xi} =  \frac{1}{4\pi}\int_{\mc H^+} dV d\Omega_2 :\op{\delta\sigma}_{AB}\op{\delta\sigma}^{AB}:,
    \label{eq:flux-op-quantum-1}
\ee
where $:\op{\delta\sigma}_{AB}\op{\delta\sigma}^{AB}:$ is normal ordered shear squared, the normal ordering being done by Hadamard subtraction w.r.t. the vacuum $\omega_0$.

We then consider the sub-algebra $\mc A_{H_A^+}$ which is generated by smeared field observables $\op{\delta\sigma}(s)$ with $Supp(s) \subseteq \mc H^+_{A}$. Then, we consider the Weyl algebra $\mathfrak A(\mc H_{A}^+, \omega_0)$, which is the closure \footnote{under strong operator topology in $\mc B(\Hilb_{\text{AdS}})$} of the algebra generated by Weyl observables $\exp(i\op{\delta\sigma}(s))$ where $\op{\delta\sigma}(s)$ is the representation of the abstract algebra element in the GNS Hilbert space $\Hilb_{\text{AdS}}$. This algebra $\mathfrak A(\mc H_{A}^+, \omega_0)$ is a Type-III von Neumann factor being a sub-region algebra of QFT due to Araki \cite{Araki} \footnote{For a comprehensive review of von Neumann algebras, factors and their types, we refer the reader to \cite{Sorcevon}}.

An extremely important fact that we shall use is that the state $\omega_0$ when restricted to $\mc H_{A}^+$ is KMS w.r.t. the horizon boost Killing field with inverse temperature $\beta = 2\pi$. This was proven by Buchholz, Florig, and Summers \cite{Summers} and we reproduce their proof in Appendix \cref{sec:KMS-proof}.

\section{Geometric modular flow and crossed product in the bulk} \label{sec:crossed-product-in-bulk}

In the previous section it has been argued that the state $|\omega_0\rangle$ is KMS w.r.t. horizon boost Killing field in $\mathfrak A(\mc H_A^+, \omega_0)$ with $\beta = 2\pi$. Now, the state $|\omega_0\rangle$ is KMS with $\beta = 1$ w.r.t. the modular flow $\gamma$ in $\mathfrak A(\mc H_A^+, \omega_0)$ generated by the modular Hamiltonian $\op H_{\omega_0} = - \log \op\Delta_{\omega_0}$ where $\op \Delta_{\omega_0}$ is the modular operator for the state $|\omega_0\rangle$ in $\mathfrak A(\mc H_A^+, \omega_0)$. \footnote{For an introduction to basic concepts of Tomita-Takesaki modular theory, especially from an algebraic QFT viewpoint, we refer the reader to the concise article \cite{Summers2} or Section-3.2 of \cite{MP}.} Then we use a powerful result in Tomita-Takesaki modular theory (theorem-2.2 of \cite{Summers2}), which states that any automorphism flow in the algebra (here Killing translations $\alpha$) under which a cyclic separating state (here $|\omega_0\rangle$) satisfies KMS property must coincide with the modular automorphism flow (here $\gamma$) of the same state in the algebra (modulo some numerical factor accounting for $\beta \neq 1$ for KMS condition w.r.t. $\alpha$). Thus, using the aforementioned result, we conclude that:
\be
    \op H_{\omega_0} = 2\pi \op F_{\xi},
    \label{eq:BW}
\ee
i.e. the modular flow of the state $|\omega_0\rangle$ in the algebra $\mathfrak A(\mc H_A^+, \omega_0)$ is the (geometric) Killing flow on the horizon $\mc H_A^+$. We shall refer to the modular automorphism group of the state $|\omega_0\rangle$ in $\mathfrak A(\mc H_A^+, \omega_0)$ as $\mathbb R_{\omega_0}$ which is isomorphic to the group of real numbers under addition.

\subsection{Dressed observables in the bulk and crossed product algebra}

Recall that the algebra  $\mathfrak A(\mc H_A^+, \omega_0)$ is a Type-III von Neumann factor and hence has no finite non-trivial trace defined on it leading to no non-zero finite density matrices (and hence no finite entropies of reduced states) in this algebra. Also, in full quantum gravity, a natural thing to wish for is general diffeomorphism invariant observables (just like in gauge theories one wishes to construct gauge-invariant observables). In our case of linearized quantum gravity, one would therefore wish to have observables invariant under Killing flow $\alpha$. However clearly the Weyl operators generating  $\mathfrak A(\mc H_A^+, \omega_0)$ are not invariant under Killing flow. Both of these issues are resolved by constructing appropriate dressed observables by extending the algebra. To do that, firstly note that the horizon flux operator $\op F_{\xi}$ belongs to the algebra $\mathfrak A(\mc H_A^+, \omega_0)$ as $F_{\xi}$ is a classical phase space observable. However the boundary charges $X$ and $C$ (defined in \cref{eq:boundary-charges-classical}) are not observables on the phase space of linearized gravity as they depend on $\delta^2g$. So they cannot be taken into the algebra $\mathfrak A(\mc H_A^+, \omega_0)$. Thus, we extend the algebra to include the boundary charges in the following way:

We consider the extended Hilbert space $\Hilb_{\text{AdS}} \otimes L^2(\mathbb R)$ with the boundary charge $\op X$ acting as position operator in $L^2(\mathbb R)$. We then extend the geometric modular flow to the auxiliary Hilbert space $L^2(\mathbb R)$ by generating it via $-\op X$. So, in the algebra $\mathfrak A(\mc H_A^+, \omega_0) \otimes \mc B(L^2(\mathbb R))$, the geometric modular flow is generated by the total Hamiltonian $\op F_{\xi} - \op X = -\op C$. Then we consider the sub-algebra of $\mathfrak A(\mc H_A^+, \omega_0) \otimes \mc B(L^2(\mathbb R))$ which commutes with the total Hamiltonian $- \op C$ \footnote{$\op C$ is also referred to as the constraint charge in the literature as the equation $F_{\xi} = X - C$ is the constraint equation on initial data on metric perturbations.} and we call it $\mathfrak A^{\text{ext}}(\mc H_A^+, \omega_0)$ which turns out to be a crossed product factor of $\mathfrak A(\mc H_A^+, \omega_0)$ with its modular automorphism group $\mathbb R_{\omega_0}$:
\be
    \mathfrak A^{\text{ext}}(\mc H_A^+, \omega_0) = \mathfrak A(\mc H_A^+, \omega_0) \rtimes \mathbb R_{\omega_0}
\ee
The fact that this total Hamiltonian invariant factor gives the crossed product algebra is dependent on the fact that the modular flow is geometric \footnote{This is because we could replace $\op F_{\xi}$ with $\op H_{\omega_0}/\beta$ and algebraically, crossed product factor is invariant under $\gamma \otimes \text{Ad }\lambda$ where $\lambda$ corresponds to the left translation unitary representation of $\mathbb R_{\omega_0}$ in $L^2(\mathbb R)$ (i.e. $\lambda(t) = \exp(-i\op p t)$ where $\op p$ is the ``momentum operator'' in $L^2(\mathbb R)$ defined as the conjugate to $\op X$ as $[\op X, \op p] = i$) and $\text{Ad } \lambda$ corresponds to the action of $\lambda$ on $\mc B(L^2(\mathbb R))$ via conjugation. Note that as the Pontrygain dual of $\mathbb R$ is $\mathbb R$ itself, so crossed product factor can be equivalently defined to be the factor invariant under $\gamma \otimes \text{Ad} \hat{\lambda}$ where $\hat{\lambda}(t) = \text{exp}(-i\op Xt)$.} and is widely known in the literature in the context of an extended algebra in the presence of an observer in de Sitter spacetime \cite{CLPW, FJLRW, Hoehn1, Hoehn2} and the dressed algebra in black hole spacetimes \cite{MP, Gautam1}. 

Now due to a theorem due to Takesaki \cite{Tak73}, the crossed product of a Type-III factor (here $\mathfrak A(\mc H_A^+, \omega_0)$) with its modular automorphism group (w.r.t. a cyclic separating state, here $|\omega_0\rangle$) is a Type-II factor (here $\mathfrak A^{\text{ext}}(\mc H_A^+, \omega_0)$). Hence the algebra $\mathfrak A^{\text{ext}}(\mc H_A^+, \omega_0)$ has a well-defined trace given by \cite{Witten22}:
\be
    \text{Tr}(\op a) = \int_{-\infty}^{\infty} dX e^{X} \langle \omega_0, X|\op a|\omega_0, X\rangle, \quad \op a \in \mathfrak A^{\text{ext}}(\mc H_A^+, \omega_0)
    \label{eq:trace-def}
\ee
where $|\omega_0, X\rangle = |\omega_0\rangle \otimes |X\rangle$.

\section{Holographic map and crossed product in the boundary} \label{sec:crossed-product-boundary}

The 2-surface $\partial \Sigma$ serves as the initial condition surface for field theory on the boundary. We shall denote the boundary field by $\phi$, which can be any tensor field which is dual to the bulk theory of linearized quantum gravity. The state $\tilde{\omega}_0$ is the boundary isometry invariant vacuum of the boundary CFT. Note that in general the QFT for $\phi$ is an interacting theory \footnote{Recall that the linearized quantum gravity sector of semi-classical limit of AdS/CFT has the bulk being described by linearized metric perturbations over solutions to Einstein's equations in the Einstein sector of supergravity effective field theory. This corresponds to large $N$ and large $\lambda$ limit of the boundary gauge theory, making it strongly coupled.} and we perform algebraic quantization of the interacting field theory \footnote{This is done perturbatively as non-perturbative quantization of interacting field theories is not yet known. For an algebraic approach to perturbative quantization of interacting theories, see \cite{RejznerPAQFT}} with $\tilde{\omega}_0$ being the interacting field vacuum. Now we do GNS construction of the field algebra $\tilde{\mc A}_{\partial\Sigma}$ w.r.t. the state $\tilde{\omega}_0$ to obtain a Hilbert space $\Hilb_{\text{CFT}}$ with a cyclic separating state $|\tilde{\omega}_0\rangle$ and a representation of the algebra $\tilde{\mc A}_{\partial \Sigma}$ in $\mc L(\Hilb_{\text{CFT}})$. We denote the Weyl algebra $\tilde{\mathfrak A}(A, \tilde{\omega}_0)$ as the closure of the algebra generated by Weyl operators with smearing functions supported in the region $A$.

Now AdS/CFT correspondence admits the existence of an isometric embedding map (referred to as the holographic map):
\be
    T: \Hilb_{\text{AdS}} \rightarrow \Hilb_{\text{CFT}}
\ee
The image space of $\Hilb_{\text{AdS}}$ under the map $T$, namely $T(\Hilb_{\text{AdS}}) \subseteq \Hilb_{\text{CFT}}$ is often referred to as the \textit{code subspace} in the literature concerning quantum error correcting properties of the holographic map \cite{Harlow, HaPPY, Ayan} and we will follow the same terminology and refer to it as $\Hilb_{\text{code}} = T(\Hilb_{AdS})$. We denote the projection to the code subspace $\Hilb_{\text{code}}$ as $P_{\text{code}} = TT^*$. 

The map $T$ has the following properties: 
\begin{enumerate}
    \item \label{prop1} It maps AdS vacuum to CFT vacuum: 
    \be 
    T|\omega_0\rangle = |\tilde{\omega}_0\rangle.
    \ee

    \item \label{prop2} It admits AdS-Rindler wedge reconstruction, i.e. 
    \be 
    T^*\tilde{\mathfrak A}(A, \tilde{\omega}_0)T = \mathfrak A(\mc H_A^+, \omega_0) \iff P_{\text{code}} \tilde{\mathfrak A}(A, \tilde{\omega}_0) P_{\text{code}} = T\mathfrak A(\mc H_A^+, \omega_0) T^*.
    \label{eq:reconstruction}
    \ee
\end{enumerate}

Now denote the algebra $P_{\text{code}} \tilde{\mathfrak A}(A, \tilde{\omega}_0) P_{\text{code}} \subset \tilde{\mathfrak A}(A, \tilde{\omega}_0)$ as the restricted CFT algebra $\tilde{\mathfrak A}_{\text{rest}}(A, \tilde{\omega}_0)$ consisting of CFT operators acting only on code subspace: 
\be
    \tilde{\mathfrak A}_{\text{rest}}(A, \tilde{\omega}_0) := P_{\text{code}} \tilde{\mathfrak A}(A, \tilde{\omega}_0) P_{\text{code}} \subseteq \mc B(\Hilb_{\text{code}}).
\ee
Note that from \cref{eq:reconstruction} we have $\tilde{\mathfrak A}_{\text{rest}}(A, \tilde{\omega}_0) = T \tilde{\mathfrak A}(\mc H_A^+, \omega_0)T^*$ and as $T: \Hilb_{\text{AdS}} \rightarrow \Hilb_{\text{code}}$ is an isometry, we conclude that $\tilde{\mathfrak A}_{\text{rest}}(A, \tilde{\omega}_0)$ is also a Type-III factor.

Now, in the algebra $\tilde{\mathfrak A}_{\text{rest}}(A, \tilde{\omega}_0)$ one has the modular operator for the state $|\tilde{\omega}_0\rangle$ which we denote by $\tilde{\op\Delta}_{\tilde{\omega}_0}$, which generates a modular flow $\tilde{\gamma}$ in $\mathfrak A_{\text{rest}}(A, \tilde{\omega}_0)$ via the modular Hamiltonian $\tilde{\op H}_{\tilde{\omega}_0} = - \log \tilde{\op \Delta}_{\tilde{\omega}_0}$. 

The map $T$ \textit{push forwards} the modular flow $\gamma$ (which was geometric) in $\mathfrak A(\mc H_A^+, \omega_0)$ to the algebra $\tilde{\mathfrak A}_{\text{rest}}(A, \tilde{\omega}_0)$ as a map $\gamma^*$ defined as follows: For any element $\tilde{\op a} \in \tilde{\mathfrak A}_{\text{rest}}(A, \tilde{\omega}_0)$, due to property \cref{prop2}, $\exists \op a \in \mathfrak A(\mc H_A^+, \omega_0)$ s.t. $\tilde{\op a} = T\op a T^*$ and $\gamma^*_t(\tilde{\op a})$ is defined as:
\be
    \gamma^*_t(\tilde{\op a}) := T \gamma_t(\op a) T^*. 
    \label{eq:gamma-star}
\ee
Now note that for any $\tilde{\op a}, \tilde{\op b} \in \tilde{\mathfrak A}(A, \tilde{\omega}_0)$ with $\tilde{\op a} = T \op a T^*$ and $\tilde{\op b} = T \op b T^*$ one has:
\be
    \langle \tilde{\omega}_0|\gamma^*_{t+i}(\tilde{\op a})\tilde{\op b}|\tilde{\gamma}_0\rangle = \langle \tilde{\omega}_0|T\gamma_{t+i}(\op a)T^*T\op b T^*|\tilde{\omega}_0\rangle = \langle  \omega_0|\gamma_{t+i}(\op a)\op b|\omega_0\rangle \\ 
    = \langle \omega_0|\op b \gamma_{t}(\op a)|\omega_0\rangle = \langle\tilde{\omega}_0|\tilde{\op b} \gamma^*_t(\tilde{\op a}) |\tilde{\omega}_0\rangle
    \label{eq:gammma-star-KMS}
\ee
where the first equality uses the definition of $\gamma^*$ as in \cref{eq:gamma-star}; the second equality uses property \cref{prop1} and the fact that $T^*T = \op 1$ on $\Hilb_{\text{AdS}}$; the third equality follows from the fact that the state KMS property of $|\omega_0\rangle$ in $\mathfrak A(\mc H_A^+, \omega_0)$ with $\beta = 1$ w.r.t. modular flow $\gamma$; and finally the fourth equality again uses property \cref{prop1} and $T^*T = \op 1$ on $\Hilb_{\text{AdS}}$. Thus from \cref{eq:gammma-star-KMS} we conclude that the state $|\tilde{\omega}_0\rangle$ is KMS (with $\beta = 1$) in the algebra $\tilde{\mathfrak A}(A, \tilde{\omega}_0)$ w.r.t. the flow $\gamma^*$. Again using theorem-2.2 of \cite{Summers2} we conclude that the flow $\gamma^*$ is same as the modular flow $\tilde{\gamma}$. Thus we have:
\be
    \tilde{\op H}_{\tilde{\omega}_0} = T \op H_{\omega_0} T^*
    \label{eq:bulk-bound-mod-flow}
\ee
We denote the modular automorphism group of the algebra $\tilde{\mathfrak A}_{\text{rest}}(A, \tilde{\omega}_0)$ w.r.t. $\tilde{\gamma}$ as $\mathbb R_{\tilde{\omega}_0}$.

Now consider the crossed product algebra: 
\be
    \tilde{\mathfrak A}^{\text{ext}}_{\text{rest}}(A, \tilde{\omega}_0) = \tilde{\mathfrak A}_{\text{rest}}(A, \tilde{\omega}_0) \rtimes \mathbb R_{\tilde{\omega}_0},
    \label{eq:boundary-cp}
\ee
which is again the subalgebra of $\tilde{\mathfrak A}_{\text{rest}}(A, \tilde{\omega}_0) \otimes \mc B(L^2(\mathbb R))$ invariant under $\frac{1}{2\pi}\tilde{\op H}_{\tilde{\omega}_0} - \op X$, where $- \op X$ again generates modular flow in $L^2(\mathbb R)$ (with $\op X$ being the position operator in $L^2(\mathbb R)$ as usual). Note that we took a factor of $1/2\pi$ in the previous line as we shall soon relate it to the crossed product in the bulk where the crossed product algebra was the invariant factor under $\op F_{\xi} - \op X$ and we had $\op F_{\xi} = \frac{1}{2\pi}\op H_{\omega_0}$ (see \cref{eq:BW}). An important point to note is that the modular flow $\tilde{\gamma}$ is not required to be geometric in $\tilde{\mathfrak A}_{\text{rest}}(A, \tilde{\omega}_0)$ and hence $\frac{1}{2\pi}\tilde{\op H}_{\tilde{\omega}_0} - \op X$ has no boundary charge interpretation. 

Now not that $T \otimes \op 1: \Hilb_{\text{AdS}} \otimes L^2(\mathbb R) \rightarrow \Hilb_{\text{code}} \otimes L^2(\mathbb R)$ is an isometry. Further, since the bulk and boundary modular flows match via \cref{eq:bulk-bound-mod-flow}, so $\frac{1}{2\pi}\tilde{\op H}_{\tilde{\omega}_0} - \op X = (T \otimes \op 1)(\op F_{\xi} - \op X)(T \otimes \op 1)^*$ and hence we conclude that:
\be
    \tilde{\mathfrak A}^{\text{ext}}_{\text{rest}}(A, \tilde{\omega}_0) = (T \otimes \op 1)(\tilde{\mathfrak A}^{\text{ext}}(\mc H^+_A, \omega_0))(T \otimes \op 1)^*.
    \label{eq:bulk-bound-crossed}
\ee

Also, being the crossed product algebra of a Type-III factor with its modular automorphism group w.r.t. some cyclic separating state, the algebra $\tilde{\mathfrak A}^{\text{ext}}_{\text{rest}}(A, \tilde{\omega}_0)$ is a Type-II factor with a well-defined trace. The trace functional can be written similar to \cref{eq:trace-def} as:
\be
    \text{Tr}(\op a) = \int_{-\infty}^{\infty} dX e^{X} \langle \omega_0, X|\op a|\omega_0, X\rangle, \quad \op a \in \tilde{\mathfrak A}^{\text{ext}}_{\text{rest}}(A, \tilde{\omega}_0)
    \label{eq:trace-def-boundary}
\ee

\subsection{Proof of JLMS condition} \label{sec:JLMS}

Now, consider some state of linearized gravity in the bulk $|\omega\rangle \in \Hilb_{\text{AdS}}$ with $\Hilb_{\text{code}} \ni |\tilde{\omega}\rangle = T|\omega\rangle$ being the corresponding CFT state in the boundary. Now consider the bulk relative modular operator $\op \Delta_{\omega|\omega_0}$ which gives the bulk relative entropy $S(\omega|\omega_0) = \langle\omega|\log \op\Delta_{\omega|\omega_0}|\omega\rangle$. \footnote{Note that the convention for defining the relative modular operators and hence relative entropy differ in literature and here we are following the convention as described in \cite{MP}.} The boundary relative modular operator $\tilde{\op \Delta}_{\tilde{\omega}|\tilde{\omega}_0}$ analogously gives the boundary relative entropy $\tilde{S}(\tilde{\omega}|\tilde{\omega}_0)$. 

Let $\op S_{\omega|\omega_0}$ be the bulk relative Tomita operator and $\tilde{\op S}_{\tilde{\omega}|\tilde{\omega}_0}$ be the corresponding boundary one. Now from the definition of relative Tomita operator we can say that for any $\tilde{\op a} \in \tilde{\mathfrak A}_{\text{rest}}(A, \tilde{\omega}_0)$:
\be
    \tilde{\op S}_{\tilde{\omega}|\tilde{\omega}_0}\tilde{\op a} |\tilde{\omega}_0\rangle = \tilde{\op a}^*|\tilde{\omega}\rangle \\
    \implies \tilde{\op S}_{\tilde{\omega}|\tilde{\omega}_0}T\op a T^* |\tilde{\omega}_0\rangle = T \op a^* T^*|\tilde{\omega}\rangle \\
    \implies T^*\tilde{\op S}_{\tilde{\omega}|\tilde{\omega}_0}T\op a |\omega_0\rangle = \op a^*|\omega\rangle \\
    \label{eq:rel-tomitas-equal}
\ee
where in the second line $\op a \in \mathfrak A(\mc H_A^+, \omega_0)$ s.t. $\tilde{\op a} = T \op a T^*$ and in the third line we used the fact that $T^*T = \op 1$ in $\Hilb_{\text{AdS}}$. Now since every operator $\op a \in \mathfrak A(\mc H_A^+, \omega_0)$ can be obtained from some $\tilde{\op a} \in \tilde{\mathfrak A}_{\text{rest}}(A, \tilde{\omega}_0)$ by $\op a = T^*\tilde{\op a} T$, so from \cref{eq:rel-tomitas-equal} we conclude that:
\be
    T^*\tilde{\op S}_{\tilde{\omega}|\tilde{\omega}_0}T = \op S_{\omega|\omega_0} \iff \tilde{\op S}_{\tilde{\omega}|\tilde{\omega}_0} = T\op S_{\omega|\omega_0}T^*.
    \label{eq:rel-tomitas-equal-final}
\ee
where the iff holds because $TT^* = \op 1$ in $\Hilb_{\text{code}}$ and $T^*T = \op 1$ in $\Hilb_{\text{AdS}}$. Now by polar decomposing both sides of \cref{eq:rel-tomitas-equal-final} we conclude that:
\be
    \tilde{\op \Delta}_{\tilde{\omega}|\tilde{\omega}_0} = T\op \Delta_{\omega|\omega_0}T^*.
    \label{eq:rel-mods-equal}
\ee

Now, taking logarithm and then expectation value w.r.t. $|\tilde{\omega}\rangle$ in both sides of \cref{eq:rel-mods-equal} we establish that:
\be
    \tilde{S}(\tilde{\omega}|\tilde{\omega}_0) = S(\omega|\omega_0),
    \label{eq:JLMS}
\ee
which is precisely the JLMS condition \cite{JLMS}. In words, it says that the relative entropy of two bulk semi-classical states in the entanglement wedge equals to the relative entropy of the dual CFT states in the code subspace CFT algebra of the corresponding boundary subregion.

\begin{remark}
    Note that in \cref{eq:rel-mods-equal} by putting $|\omega\rangle = |\omega_0\rangle$ and taking negative logarithm both sides, one can re-derive \cref{eq:bulk-bound-mod-flow}. Hence this method serves as a second way to prove that the modular flow in bulk induces modular flow in code subspace in the boundary. 
\end{remark}

\begin{remark}
    Note that the proof of JLMS condition didn't require the modular flow to be geometric in the bulk (which although it is for our case). Crossed product construction is also not needed to prove it. It only required the existence of a holographic map satisfying the properties \cref{prop1} and \cref{prop2}
\end{remark}

\section{Coherent states of gravity in the bulk and proof of HRT formula} \label{sec:HRT-proof}

Now consider the initial data of linearized (i.e. first-order) metric perturbations $h_{AB}$ to be supported in $\mc H^+_A$. In the quantum field theory, this metric perturbation can be represented by a coherent state defined as follows. For any perturbation $h_{AB}$ on $\mc H^+_A$, define the unitary
\be
    \op{U} \defn \exp(-i\op{\delta\sigma}(h)/16\pi^2).
\ee
The (algebraic) coherent state on $\mc H^+_A$ due to the metric perturbation $h_{AB}$ is then given by
\be
    \omega_h(\op{a}) \defn \omega_0(\op{U^*aU}) \quad \forall \op{a} \in \mathfrak A(\mc H^+_A, \omega_0).
    \label{eq:def-coherent-state}
\ee
From the above definition it follows that the 1-point function of this coherent state is $\omega_h(\op{\delta\sigma}(s)) = \int dV d\Omega_2~ \delta\sigma_{AB} s^{AB}$ where $\delta\sigma_{AB} = \frac{1}{2}\partial_V h_{AB}$ is the shear of the perturbation and hence, this coherent state is indeed the perturbed state of the AdS spacetime.

Now we extend the state to entire $\mc H^+$ by defining the state $|\omega_h\rangle \in \Hilb_{\text{AdS}}$ to be the unique state in the natural cone $P^{\#}$ of $(\mathfrak A(\mc H_A^+, \omega_0), |\omega_0\rangle)$ in $\Hilb_{\text{AdS}}$ as:
\be
    |\omega_h\rangle = \op U j_{\omega_0}(\op U)|\omega_0\rangle
\ee
where $j_{\omega_0}: \mathfrak A(\mc H_A^+, \omega_0) \rightarrow \mathfrak A(\mc H_A^+, \omega_0)'$ defined by $j_{\omega_0}(\op a) = \op J_{\omega_0} \op a \op J_{\omega_0}$. In the preceding line, $\mathfrak A(\mc H_A^+, \omega_0)'$ is the commutant of $\mathfrak A(\mc H_A^+, \omega_0)$ (and is equal to $\mathfrak A(\mc H_{A^c}^+, \omega_0)$ by Haag duality) and $\op J_{\omega_0}$ is the modular conjugation operator for the state $|\omega_0\rangle$ in $\mathfrak A(\mc H_A^+, \omega_0)$. Note that taking the state $|\omega_h\rangle$ in the natural cone is just a standard modular theoretic convention (which simplifies many computations) and does not change anything. In fact as one might expect that a more ``natural'' candidate for $|\omega_h\rangle$ should have been just $\op U|\omega_0\rangle$. However in the Appendix \cref{sec:appA}, it has been shown that they give exactly same result for the final crossed product entropy (see \cref{eq:def-entropy}). 

Corresponding to the coherent state \(\ket{\omega_h}\), we define a ``classical-quantum state'' \(\ket{\underline{\omega}_h}\) in the extended Hilbert space $\Hilb_{\text{AdS}} \otimes L^2(\mathbb{R})$ by
\be 
    \ket{\underline{\omega}_h} \defn \int_{\mathbb{R}}dX f(X) \ket{ \omega_h} \otimes \ket{X},
\ee
where $f \in L^2(\mathbb{R})$ which can be interpreted as the ``wave function of the boundary charge at $\zeta_R^+$''.

Since the extended algebra $\mf A^{\text{ext}}(\mc H_A^+, \omega_0)$ is a Type-II factor, there exists a (renormalized) density matrix $\op \rho_{\underline{\omega}_h} \in \mf A^{\text{ext}}(\mc H_A^+, \omega_0)$ corresponding to the state $\ket{\underline{\omega}_h}$ such that
\be
    \text{Tr} (\op \rho_{\underline{\omega}_h}\op a) = \langle\underline{\omega}_h|\op a|\underline{\omega}_h\rangle \quad \forall \op a \in \mf A^{\text{ext}}(\mc H_A^+, \omega_0) \eqsp \text{Tr} (\op\rho_{\underline{\omega}_h}) = 1,
    \label{eq:density-matrix-def}
\ee
where the trace functional $\text{Tr}$ is defined on $\mf A^{\text{ext}}(\mc H_A^+, \omega_0)$ in \cref{eq:trace-def}.
This density matrix $\op\rho_{\underline{\omega}_h}$ serves as an analogue of the reduced density matrix of the full semi-classical coherent state on the entire initial data surface $\mc H^+$ to the AdS-Rindler horizon $\mc H^+_A$ corresponding to region $A$. Note that this reduced density matrix is not just a reduced QFT state to $\mc H_A^+$ (which is undefined) but a joint reduced state of the QFT and boundary charge at $\zeta_R^+$ to the horizon $\mc H^+_A$.

Furthermore the trace \cref{eq:trace-def} allows one to compute the von Neumann entropy of the density matrix $\op \rho_{\underline{\omega}_h}$ by:
\be
    S(\op \rho_{\underline{\omega}_h}) = -\text{Tr}(\op \rho_{\underline{\omega}_h} \log \op \rho_{\underline{\omega}_h}).
    \label{eq:def-entropy}
\ee
This computation has been done for classical quantum states (e.g. see Appenix A of \cite{MP}) for ``slowly-varying wave function" $f(X)$ (roughly speaking, the observer wave function's Fourier transform is sharply peaked at zero momentum). \footnote{It can be shown that the difference between the actual von Neumann entropy and the entropy obtained by taking slowly varying wave-function approximation is bounded between $0$ and a function depending on the observer wave function $f$. For details, see Lemma 2 of Appendix B in \cite{GautamCov}.} The result that one obtains is:
\be
    S(\op \rho_{\underline{\omega}_h}) = -S(\omega_h|\omega_0) + 2\pi\langle \op X\rangle_{\underline{\omega}_h} + S(f), 
    \label{eq:bulk-entropy}
\ee
where $S(f) = -\int_{\mathbb{R}}dX|f(X)|^2\log|f(X)|^2$. The relative entropy between a coherent state and a vacuum state has been is computed in \cite{HI, GautamInfo}. First we note that \cite{HI} computed $S(\omega_0|\omega_h)$ instead of $S(\omega_h|\omega_0)$. However for $\ket{ \omega_h }$ in the natural cone, it can be shown that they are equal. Secondly, \cite{HI} uses the modular operator in the von Neumann algebra generated by Weyl operators supported in the future of some cut with \(V = V_0\), while \cite{GautamInfo} uses the past of some cut. Thus, their expression for relative entropy includes the affine time \(V_0\) of this cut. The corresponding modular flow then corresponds to a ``dilation'' which keeps $V_0$ fixed. In our case, the relevant algebra is the one supported on \(\mc H^+_A\) and \(V_0 = 0\). Thus the relative entropy in our case is
\be\label{relentropy}
    S(\omega_h|\omega_0) = 2\pi F_{\xi}[\mc H_A^+] = \frac{1}{2}\int_{\mc H_A^+}dV d\Omega_2 V(\delta\sigma_h)^2
, \ee
where $\delta\sigma_h = \frac{1}{2}\partial_Vh$ is the perturbed shear at $\mc H^+$ corresponding to $h_{AB}$ and hence $F_{\xi}[\mc H_A^+]$ is the classical radiation flux of the linearized perturbation falling through $\mc H_A^+$. 

The boundary-charge wave function $f(X)$ can be assumed to be peaked at the classical value of the boundary charge $X$. If not so, then the shift will just contribute to a constant number. In any case, gathering all the $\omega_h$ state-dependent terms of $S(\op \rho_{\underline{\omega}_h})$ we have: 
\be
    S(\op \rho_{\underline{\omega}_h}) = 2\pi(X - F_{\xi}[\mc H_A^+]) + (\omega_h-\text{independent terms}) \\ = 2\pi \int_{\Gamma}\delta^2Q_{\xi} + (\omega_h-\text{independent terms}),
    \label{eq:bulk-entropy-2}
\ee
where the first equality follows from \cref{eq:bulk-entropy} and for the second equality we used the fact that $F_{\xi}[\mc H_A^+] = \int_{\mc H_A^+}d(\delta^2Q_{\xi} - \xi\cdot \delta\theta(g, \delta g)) = X - \int_{\Gamma} \delta^2Q_{\xi}$ as $\xi^a = 0$ on $\Gamma$. Now, the Noether charge of gravity in the case of general relativity at the bifurcation surface of a Killing horizon evaluates to \cite{Iyer-Wald, HWZ}:
\be
    \int_{\Gamma}\delta^2Q_{\xi} = \frac{1}{8\pi}\delta^2 \text{Ar}[\Gamma_0],
    \label{eq:area}
\ee
where $\delta^2\text{Ar}[\Gamma_0]$ is the (second-order) perturbed area of the surface $\Gamma_0$. Thus putting \cref{eq:area} in \cref{eq:bulk-entropy-2} we obtain:
\be
    S(\op \rho_{\underline{\omega}_h}) = \frac{1}{4}\delta^2\text{Ar}[\Gamma_0] + (\omega_h-\text{independent terms}).
    \label{eq:final-bulk-entropy}
\ee

Now action of $T \otimes \op 1$ on $|\underline{\omega}_h\rangle$ produces a state $|\tilde{\underline{\omega}}_h\rangle \in \Hilb_{\text{code}}\otimes L^2(\mathbb R)$:
\be 
    \ket{\tilde{\underline{\omega}}_h} \defn (T \otimes \op 1)|\underline{\omega}_h\rangle = \int_{\mathbb{R}}dX f(X) T\ket{ \omega_h} \otimes \ket{X},
\ee
and we shall further denote the state $T|\omega_h\rangle$ as $|\tilde{\omega}_h\rangle$; hence $\ket{\tilde{\underline{\omega}}_h}$ is the corresponding classical-quantum state in the CFT. Now one can compute the corresponding reduced density matrix $\op \rho_{\tilde{\underline{\omega}}_h} \in \mathfrak A_{\text{rest}}^{\text{ext}}(A, \tilde{\omega}_0)$ using the trace in \cref{eq:trace-def-boundary} by employing similar prescription as \cref{eq:density-matrix-def} for the $\mathfrak A_{\text{rest}}^{\text{ext}}(A, \tilde{\omega}_0)$ algebra. The entropy of the density matrix $\op \rho_{\tilde{\underline{\omega}}_h}$ can be evaluated in exactly similar way and one obtains: 
\be
    S(\op \rho_{\tilde{\underline{\omega}}_h}) = -S(\tilde{\omega}_h|\tilde{\omega}_0) + 2\pi \langle \op X\rangle_{\tilde{\underline{\omega}}_h} + S(f)
\ee
Now by using JLMS condition (see \cref{eq:JLMS}) we can say:
\be
    S(\op \rho_{\tilde{\underline{\omega}}_h}) = -S(\omega_h|\omega_0) + 2\pi \langle \op X\rangle_{\tilde{\underline{\omega}}_h} + S(f) = S(\op \rho_{\underline{\omega}_h})
\ee
and hence by using \cref{eq:final-bulk-entropy} the entropy of the reduced state in boundary CFT in the crossed product algebra can be written as:
\be
    S(\op \rho_{\tilde{\underline{\omega}}_h}) = \frac{1}{4}\delta^2\text{Ar}[\Gamma_0] + (\tilde{\omega}_h-\text{independent terms}).
    \label{eq:HRT-proved}
\ee

Now recall from \cref{eq:principal-eq-area} that $\delta^2\text{Ar}[\Gamma_0] = (\text{Ar}[\Gamma_{\lambda}, g_{ab}(\lambda)] - \text{Ar}[\Gamma_0, g_{ab}^{(0)}]) + \mc O(\lambda^3)$. Hence the R.H.S. of \cref{eq:HRT-proved} is a quarter of the vacuum subtracted area of the HRT surface of the state $\omega_h$ at leading order (here second order) in $\lambda$. This shows that at leading order in $\lambda$, the state dependent part of the Type-II entropy of the crossed product state $\underline{\tilde{\omega}}_h$ is given by vaccum subtracted area of HRT surface, which is precisely the R.H.S. of \cref{eq:intro-main} and thus we have proven what we advertised in the introduction. To reiterate: the vacuum subtracted entropy (which a priori is the subtraction of two ill-defined UV-divergent entropies) should be interpreted as the entropy of the crossed product state (which is well-defined) which in turn is given by the vacuum subtracted area of the HRT surface (at leading order in perturbation):
\be
    \Delta S(\omega_h, A) = S(\op \rho_{\tilde{\underline{\omega}}_h}) = \frac{1}{4}(\text{Ar}[\Gamma_{\lambda}, g_{ab}(\lambda)] - \text{Ar}[\Gamma_0, g_{ab}^{(0)}]) \quad \text{(all equalities at } \mc O(\lambda^2))
\ee

\section{Limitations and future directions} \label{sec:outro}
 
We conclude this paper by mentioning about the implicit limitations in our constructions and proofs. We shall also point to some future directions of work that can be pursued along these lines. 

The proof of JLMS condition in \cref{sec:JLMS} is completely general. It didn't require anything except the existence of the holographic map satisfying properties \cref{prop1} and \cref{prop2}. The proof was done at the Type-III algebra level and didn't require modular crossed product construction. 

On the other hand, the proof of HRT formula in \cref{sec:HRT-proof} using modular crossed product construction has certain limitations. Firstly, it works only when the CFT state $|\tilde{\omega}_h\rangle \in \Hilb_{\text{code}}$ is dual to a coherent state of gravity $|\omega_h\rangle \in \Hilb_{\text{AdS}}$ in the bulk with the linearized metric perturbations supported in the causal wedge $\mc W(A)$ (and elsewhere where it propagates through reflecting boundary conditions) where $A$ is the boundary subregion where the entropy of the CFT state is to be computed. So, when the CFT state is not dual to a coherent state in the bulk (e.g. dual to a 1-graviton state in the bulk), our construction will not work. Also, when the entropy of the dual CFT state is being computed in some other boundary subregion $A'$ such that bulk excitations are not fully supported in $\mc W(A')$ \footnote{modulo reflections due to boundary conditions} then our construction won't work. Secondly, it requires $A$ to be a connected ball shaped region. This excludes many scenarios: for example, when $A$ is a union of multiple disconnected ball-shaped regions in the boundary, then the HRT surface might not be the sum of individual HRT surfaces of the ball-shaped regions. This is because there can now exist multiple extremal surfaces and HRT proposal says that in that case one should take the surface with minimum area. This case cannot be probed by our construction which relies on one ball-shaped boundary region $A$ and also implicitly assumes the uniqueness of extremal surface anchored to $A$ even in the perturbed spacetime. 

In spite of the limitations mentioned in the previous paragraph, our construction establishes HRT formula in a very natural regime in the semi-classical context: given a ball-shaped boundary $A$, for coherent excitations in bulk supported in the casual wedge $\mc W(A)$ \footnote{initially in $\mc W(A)$ and then goes to other areas due to reflecting boundary conditions}, the (vacuum subtracted) entropy of the dual CFT state in the boundary subregion algebra of $A$ is given by the HRT formula. This is interesting in many physically reasonable scenarios, for example to compute how entanglement of a fixed boundary region $A$ with $A^c$ change as one has localized gravitational waves in $\mc W(A)$ compared to vacuum. Also this is precisely the scenario where bulk gravity was ``deduced'' from entanglement of boundary CFT in \cite{FaulknerGrav} in the following way: taking the dual CFT state to a bulk coherent state, the first law of entanglement entropy \cite{Firstlaw} along with HRT formula concludes that the perturbations defining the bulk coherent state must satisfy linearized Einstein's equations. \footnote{In \cite{FaulknerGrav} it was also showed that for the case where the bulk gravity is some general diffeomorphism invariant theory of gravity, then if the HRT formula is replaced by a general Wald-entropy term (i.e. integral of Noether charge $Q_{\xi}$ of the gravity theory at the AdS-Rindler bifurcation surface $\Gamma_0$), then by using first law of entanglement entropy in the boundary, one can show that the metric perturbations denoting the bulk state must satisfy linearized equation of motion of the respective gravity theory.}

Also note that our construction works for perturbations over pure AdS. This excludes some interesting cases like perturbations over eternal AdS-Schwarzschild black hole spacetimes (where the dual CFT state is a thermofield double state on two copies of the CFT at two asymptotically AdS boundaries of the maximally extended AdS-Schwarzschild spacetime \cite{Maldablack}). We believe that this work can be extended to semi-classical holography on AdS black hole backgrounds. 

Finally note that the initial data of metric perturbations on $\mc H^+$ was assumed to be memory-less. \footnote{That is, $h_{AB}(\zeta_R^+) = 0$ was assumed where $h_{AB}$ was initial data of metric perturbations. Note that $h_{AB}$ being supported only on $\mc H^+_A$, $h_{AB}(V, x^A) = 0$ $\forall V \leq 0$ and hence the memory observable on $\mc H^+$, $\Delta_{AB} = h_{AB}(\zeta^+_R) - h_{AB}(\zeta^-_L) = h_{AB}(\zeta_R^+)  = 0$.} It is now well understood that such low frequency radiation plays a crucial role in the quantum theory at Killing horizons \cite{Danielson2023, Gralla2023}. The formalism of Tomita-Takesaki theory has recently been generalized to include such soft modes in \cite{GautamInfo}. Using this formalism, we expect that our analysis can be extended to account for such linearized perturbations with non-trivial memory on the AdS-Rindler horizon.

\acknowledgements

I thank Kartik Prabhu for many insightful discussions in the course of this work. I thank Ayan Mukhopadhyay and Kartik Prabhu for comments on an initial draft of this manuscript. I would also like to thank Raman Research Institute's Visiting Student Program for support.

\appendix 

\section{Proof of KMS nature of AdS vacuum in AdS-Rindler wedge} \label{sec:KMS-proof}

The state $\omega_0$ when restricted to $\mc H_{A}^+$ is KMS w.r.t. the horizon boost Killing field with inverse temperature $\beta = 2\pi$. We shall prove this by assuming the following two facts:
\begin{enumerate}
    \item \label{assump1} Linearized quantum gravity has (at least) one AdS-isometry invariant algebraic state (vacuum), which we consider as $\omega_0$. Furthermore, for the representative of this state in the GNS Hilbert space, the assumption \cref{assump2} holds. 

    \item \label{assump2} The state $|\omega_0\rangle$ is passive in $\mathfrak A(\mc H_A^+, \omega_0)$. 
\end{enumerate}

The notion of passivity is described in \cite{Pusz, Summers}. In assumption \cref{assump2}, passivity essentially means that if $\op F_{\xi}$ is the boost Hamiltonian (the flux operator), and $\op O \in \mathfrak A(\mc H_A^+, \omega_0)$ be any operator in the algebra, then:
\be
    \langle \omega_0| \op F_{\xi} |\omega_0\rangle \leq  \langle \omega_0| \op O^*\op F_{\xi} \op O |\omega_0\rangle,
\ee
i.e. it is not possible to extract energy from the vacuum state and this assumed property of the vacuum is referred to as the mathematical expression of the second law of thermodynamics in the context of quantum field theory \cite{Summers}. 

The proof of KMS property of AdS vaccum in AdS-Rindler wedge employs an important theorem due to Pusz and Woronowicz (Theorem-1.3 of \cite{Pusz}) and our proof follows similar arguments as Buchholz, Florig, and Summers in \cite{Summers} where it was shown that AdS vacuum is KMS in AdS-Rindler wedge. We reproduce similar lines of arguments as \cite{Summers} in this subsection. 

The important theorem of \cite{Pusz} which is central to our proof is as follows: Let $\mathfrak A$ is a $C^*$-algebra with an algebraic state $\psi_0$ with a one-parameter group of automorphisms $\alpha$ acting on $\mathfrak A$. \footnote{The pair $(\mathfrak A, \alpha)$ is often called a $C^*$ \textit{dynamical system} in the literature.} Suppose $G$ is a locally compact amenable group of automorphisms acting on $\mathfrak A$ with the group action commuting with $\alpha$. Suppose the state $\psi_0$ is passive in $\mathfrak A$ and weakly clustering w.r.t. $G$-action. Then the state $\psi_0$ is either a ground state w.r.t. $\alpha$ (i.e. $\psi_0(H_{\alpha}) = 0$ with $H_{\alpha} \geq 0$ where $H_{\alpha}$ is the generator of the $\alpha$-automorphisms) or it is KMS w.r.t. $\alpha$ with some positive inverse temperature $\beta$. 

The notions of amenable group and weak clustering under a group action is explained in Appendix \cref{sec:appB}.

Now here the algebra to be considered is the von Neumann factor $\mathfrak A(\mc H_A^+, \omega_0)$ with the state being $|\omega_0\rangle$ and the one-parameter family of automorphisms being the Rindler boost $\alpha$ on the horizon. By assumption \cref{assump2} it is already passive in $\mathfrak A(\mc H_A^+, \omega_0)$. Now take the locally compact amenable group $G$ to be real numbers under addition with it acting on $\mathfrak A(\mc H_A^+, \omega_0)$ via the Rindler boost $\alpha$. The $G$-action being $\alpha$ itself trivially commutes with $\alpha$ automorphism. The $\alpha$ automorphism (or equivalently the $G$-action) acts on $\mathfrak A(\mc H_A^+, \omega_0)$ via conjugation with an unitary representation of $G = \mathbb R$ in $\Hilb_{\text{AdS}}$, namely the Killing time translation operator $\op U(t) = \exp(-i\op F_{\xi}t)$:
\be
    \alpha_t(\op O) = \op U(t)^* \op O \op U(t); \hspace{10pt} \op O \in \mathfrak A(\mc H_A^+, \omega_0)
    \label{eq:killing-time-translation}
\ee 
We denote the invariant mean on the amenable group $G = \mathbb R$ by $m: L^{\infty}(G; \mathbb C) \to \mathbb C$. Now:
\be
    m[\langle \omega_o| \op A \alpha_t(\op B)|\omega_0\rangle] = m[\langle \omega_o| \op A \op U(t)^* \op B \op U(t)|\omega_0\rangle] = m[\langle \omega_0| \op A \op U(t)^* \op B |\omega_0\rangle], 
    \label{eq:clustering-1}
\ee 
where in the first equality we used \cref{eq:killing-time-translation} and in the second equality we employed the fact that the state $|\omega_0\rangle$ is AdS-isometry invariant (and hence $\op U(t)|\omega_0\rangle = |\omega_0\rangle \iff \op F_{\xi}|\omega_0\rangle = 0$). 
Now, by mean ergodic theorem (see Appendix \cref{sec:mean-ergodic} and more specifically \cref{remlast}), we have:
\be
    m[\langle \omega_0| \op A \op U(t)^* \op B |\omega_0\rangle] = \langle \omega_0| \op A \op P_{\text{inv}} \op B |\omega_0\rangle,
    \label{eq:use-mean-ergodic}
\ee
where $\op P_{\text{inv}}$ is the projector to the subspace of $\Hilb_{\text{AdS}}$ invariant under the action of representation of $G$ on $\Hilb_{\text{AdS}}$ (i.e. under Rindler boost flow $\op U(t)$). Now note that the flux operator generates Rindler boost on the horizon and $|\omega_0\rangle$ is the unique Rindler boost invariant state. So we have $\op P_{\text{inv}} = |\omega_0\rangle\langle \omega_0|$. Thus, from \cref{eq:clustering-1}, we have:
\be
    m[\langle \omega_o| \op A \alpha_t(\op B)|\omega_0\rangle] = \langle\omega_0|\op A|\omega_0\rangle\langle \omega_0|\op B|\omega_0\rangle
     \label{eq:clustering}
\ee
Hence the state $|\omega_0\rangle$ is weakly clustering under Rindler boost action. Thus using the theorem from \cite{Pusz} stated earlier, we conclude that the state $|\omega_0\rangle$ is KMS with some $\beta > 0$. Note that $|\omega_0\rangle$ cannot be a ground state as although $|\omega_0\rangle$ has zero energy w.r.t. the boost Hamiltonian $\op F_{\xi}$, but $\op F_{\xi}$ is not a positive semi-definite operator. Furthermore it was argued in \cite{Summers} that the inverse temperature for AdS vacuum in AdS-Rindler wedge w.r.t. horizon boost is $\beta = 2\pi$.

\section{Crossed product von Neumann entropy is purification independent}
\label{sec:appA}

Say $\omega$ is an algebraic state defined on a von Neumann algebra $\mathfrak A \in \mc B(\Hilb)$. Now the state can be defined as a vector in $\Hilb$ and this process is called \textit{purification}. Clearly the purification process is not unique and say $|\omega_1\rangle, |\omega_2\rangle \in \Hilb$ are two such purifications, i.e.
\be
    \langle\omega_1|\op a|\omega_1\rangle = \langle \omega_2|\op a|\omega_2 \rangle = \omega(\op a) \quad \forall \op a \in \mathfrak A. 
\ee
Now suppose $\mathfrak A$ be a Type-III factor with a cyclic separating state $|\omega_0\rangle \in \Hilb$ and its crossed product algebra with modular automorphism group of $|\omega_0\rangle$: $\mathfrak A \rtimes \mathbb R$ is a Type-II factor with a trace functional $\text{Tr}$ defined on it. 

Now consider the following states in $\Hilb \otimes L^2(\mathbb R)$: 
\be
    |\hat{\omega}_1\rangle = |\omega_1\rangle \otimes |f\rangle; \\
    |\hat{\omega}_2\rangle = |\omega_2\rangle \otimes |f\rangle,
\ee
where $|f\rangle = \int_{\mathbb R}dX f(X)|X\rangle \in L^2(\mathbb R)$. Now clearly we have:
\be
    \langle\hat{\omega}_1|\op a|\hat{\omega}_1\rangle = \langle \hat{\omega}_2|\op a|\hat{\omega}_2 \rangle \quad \forall \op a \in \mathfrak A \rtimes \mathbb R. 
    \label{eq:appA}
\ee
Now define $\op \rho_{\hat{\omega}_i} \in \mathfrak A$ (for $i = 1,2$) such that:
\be
    \text{Tr}(\op \rho_{\hat{\omega}_i}\op a) = \langle \hat{\omega}_i|\op a|\hat{\omega}_i\rangle; \quad \forall \op a \in \mathfrak A \rtimes \mathbb R \quad \text{ and } \quad \text{Tr}(\op \rho_{\hat{\omega}_i}) = 1.
    \label{eq:appA-2}
\ee
From the definition in \cref{eq:appA-2} and using \cref{eq:appA} we clearly have:
\be
    \op \rho_{\hat{\omega}_1} = \op \rho_{\hat{\omega}_2},
    \label{eq:appA-same}
\ee
and hence:
\be
    S(\op \rho_{\hat{\omega}_1}) = S(\op \rho_{\hat{\omega}_2}).
    \label{eq:appA-same-2}
\ee
Also note that other algebra $\mathfrak A$ intrinsic notions like relative entropy w.r.t. vacuum are trivially same: $S(\omega_1|\omega_0) = S(\omega_2|\omega_0)$.

Now for our case where $\mathfrak A = \mathfrak A(\mc H_A^+, \omega_0)$, take $|\omega_1\rangle = \op Uj_{\omega_0}(\op U)|\omega_0\rangle$ and $|\omega_2\rangle = \op U|\omega_0\rangle$. Now for any $\op a \in \mathfrak A(\mc H_A^+, \omega_0)$: 
\be
    \braket{ \omega_1 | \op a | \omega_1 } &= \braket{ \omega_0 | j_{\omega_0}(\op U^*)\op U^* \op a \op U j_{\omega_0}(\op U) | \omega_0 } \\
    &= \braket{ \omega_0 | j_{\omega_0}(\op U^*)j_{\omega_0}(\op U)\op U^* \op a \op U \op |\omega_0 } \\
    &= \braket{ \omega_0|\op U^* \op a \op U \op |\omega_0 } \\
    &= \braket{ \omega_2 | \op a | \omega_2 }
, \ee
where in the first line we used the fact that $\op J_{\omega_0}^* = \op J_{\omega_0}$ and hence $j_{\omega_0}(\op U)^* = j_{\omega_0}(\op U^*)$. While going to the second line, we used the fact that $j_{\omega_0}(\op U)$ is in the commutant algebra and hence we commuted $j_{\omega_0}(\op U) \in \mf A(\mc H_{A^c}^+, \omega_0)$ across operators $\op U, \op U^*, \op a \in \mf A(\mc H_A, \omega_0)$. Then in the third equality we used $\op J_{\omega_0}^2 = \op 1$ and hence $j_{\omega_0}(\op U^*) j_{\omega_0}(\op U) = \op 1$. So $|\omega_1\rangle$ and $|\omega_2\rangle$ are purifications of the same coherent state \cref{eq:def-coherent-state}. 

One important thing to note is that the state $|\omega_2\rangle$ behaves like vacuum $|\omega_0\rangle$ in $\mc H_{A^c}^+$, i.e.: 
\be
\braket{ \omega_2| \op a |\omega_2} = \braket{ \omega_0| \op a |\omega_0} \text{ for } \op a \in \mf A(\mc H_{A^c}^+, \omega_0),
\ee
and hence behaves more akin to a state which corresponds to perturbations supported only in $\mc H_A^+$. On the other hand the natural cone purified state $|\omega_1\rangle$ behaves very different from $|\omega_0\rangle$ in $\mc H_{A^c}^+$. In fact for $\op a \in \mf A(\mc H_{A^c}^+, \omega_0)$:
\be
    \braket{\omega_1| \op a |\omega_1} &= \braket{\omega_0|\op U^* j_{\omega_0}(\op U^*)\op a j_{\omega_0}(\op U) \op U|\omega_0} \\
    &= \braket{\omega_0|j_{\omega_0}(\op U^*)\op a j_{\omega_0}(\op U) |\omega_0} 
\ee
So the natural cone purified state $|\omega_1\rangle$ state induces some non-trivial perturbations even in $\mc H_{A^c}^+$ (unlike $|\omega_2\rangle$ which behaves like vacuum $|\omega_0\rangle$ in $\mc H_{A^c}^+$). However, although the state $|\omega_1\rangle$ looks very different from $|\omega_2\rangle$ in $\mc H_{A^c}^+$, still because of \cref{eq:appA-same} and \cref{eq:appA-same-2} we have their entropies in the crossed product algebra to match in $\mc H_A^+$.

\section{Amenable groups, weak clustering and mean ergodic theorem} \label{sec:appB}

\subsection{Amenable groups}

Amenability of a group $G$ formalizes the notion of existence of a measure on $G$ which is invariant under group actions and the measure of the entire group $G$ is finite. Groups having such a measure are called amenable. In this subsection, we formalize this definition and also give an equivalent definition in terms of existence of group action invariant means on $L^{\infty}$ functions on $G$. Note that everywhere in this subsection, left translation (resp. invariance) can be replaced by right translation (resp. invariance). For more detailed exposition refer to \cite{Garrido} or the classic text \cite{Paterson}.

Let $G$ be a locally compact group. If $G$ has a finitely additive probability measure $\mu$ (i.e. $\exists \mu : \mathcal{B}(G) \to [0, \infty)$ where $\mathcal B(G)$ is the set of Borel sets in $G$ such that $\mu(G) = 1$ and for a finite union of disjoint Borel measurable sets $A_i$ in $G$, $\mu(\bigsqcup_{i=1}^{n}A_i) = \sum_{i = 1}^n\mu(A_i)$) which is left translation invariant (i.e. $\mu(g\cdot A) = \mu(A)$ $\forall g \in G$ and $\forall A \in \mathcal B(G)$), then $G$ is said to be amenable. 

Very importantly note that in the above definition for the group $G$ to be amenable we need a finitely additive probability measure $\mu$ on $G$ (which is much less restrictive than a countable additive probability measure). 

\begin{remark}
    Note that all compact groups are amenable. This can be seen as follows: A locally compact group $G$ has a left-invariant countably (and hence finitely) additive measure called Haar measure $\mu_H$. Now if $G$ is compact, then $\mu_H(G) < \infty$ and hence once can get the required left-invariant probability measure $\mu = \mu_H/\mu_H(G)$.
\end{remark}

\begin{remark}
    $\mathbb R$ under addition is amenable (and this is very important as the group of Killing isometries is $\mathbb R$ under addition). In fact, historically speaking, Lebesgue motivated the idea of amenability first when he tried to get a probability measure on $\mathbb R$ which is finitely additive and translation-invariant. Note that the standard Lebesgue measure on $\mathbb R$ is countably additive but not a probability measure. In fact there cannot exist a countably additive probability measure on $\mathbb R$, but there does exist a finitely additive probability measure, thus making $\mathbb R$ amenable.
\end{remark}

Now given an amenable group $G$ with amenable measure (i.e. finitely additive left-invariant probability measure) $\mu$, one can construct a mean functional on bounded complex valued functions on $G$, denoted by $L^{\infty}(G; \mathbb C)$ as $L^{\infty}(G; \mathbb C) \ni f = u + iv \to \int u d\mu + i \int v d\mu$ such that: 
\begin{itemize}
    \item $\int f d\mu \geq 0$ if $f$ is real-valued and $f(g) \geq 0$ $\forall g \in G$;
    \item $\int 1_G d\mu = 1$ where $1_G$ is identity function on $G$;
    \item $\int f_g d\mu = \int f d\mu$ where $f_g(h) := f(g\cdot h)$ for any choice of $g \in G$
\end{itemize}
Such a linear functional is called a left invariant mean on $G$. 

\begin{remark}
    Note that a group has an amenable measure iff it has a left-invariant mean. This can be seen as follows: Existence of an amenable measure implies existence of a left invariant mean which has already been shown above. To go the other way, given a left-invariant mean $m: L^{\infty}(G; \mathbb C) \to \mathbb C$, one can get an amenable measure by defining $\mu(A) = m(1_A)$ where $1_A$ is the indicator function on the Borel set $A \subseteq G$.
\end{remark}

\subsection{Weak clustering under group action}

Let $G$ be an amenable group acting on the algebra $\mathfrak A$ which has a state $\omega$ defined on it. Say the group action is denoted by $\alpha$ and so:
\be
    \alpha: G \to \text{Aut}(\mathfrak A)
\ee
and we denote for some $g \in G$, $\alpha_g: \mathfrak A \to \mathfrak A$ as an automorphism on $\mathfrak A$ due to the action $\alpha$. Now being an amenable group, $G$ has an invariant mean defined on $L^{\infty}(G; \mathbb C)$ functions $m: L^{\infty}(G; \mathbb C) \to \mathbb C$. Now the state $\omega$ is said to be weakly clustering under the $G$-action $\alpha$ if:
\be
    m[\omega(\op a \alpha_g(\op b))] = \omega(\op a)\omega(\op b), \quad \forall \op a, \op b \in \mathfrak A.
    \label{eq:weak-clustering-def}
\ee
Essentially it means that the 2-point correlations among the algebra elements are ``clustered'', i.e. when one makes one of the algebra elements to move around in the algebra by the group action and then take the average, the correlations are ``washed away'' and the connected 2-point correlation vanishes.


\subsection{Mean ergodic theorem for amenable group representations}
\label{sec:mean-ergodic}

Say $G$ is an amenable group and it has a strongly continuous unitary representation $U: G \to \mathcal U(\Hilb)$ on a Hilbert space $\Hilb$ and we denote the unitary corresponding to $g \in G$ as $\op U(g)$. Let us denote the invariant subspace of $\Hilb$ under $U$-actions to be:
\be
    \text{Inv}(\Hilb, U) \defn \{|\psi\rangle \in \Hilb | \op U(g)|\psi\rangle = |\psi\rangle \forall g \in G\},
\ee
and the projector to the subspace $\text{Inv}(\Hilb, U)$ be $\op P_{\text{inv}}$. Then by denoting the amenable mean on $G$ to be $m$, for any $|\phi_1\rangle, |\phi_2\rangle \in \Hilb$, mean ergodic theorem states that:
\be
    m[\braket { \phi_1| \op U(g)| \phi_2}] = \braket { \phi_1| \op P_{\text{inv}} |\phi_2}.
    \label{eq:mean-ergodic-def}
\ee

\begin{remark} \label{remlast}
    \underline{Proof of \cref{eq:use-mean-ergodic}}: L.H.S. of \cref{eq:use-mean-ergodic} can be written as:
    \be
        m[\langle \omega_0| \op A \op U(t)^* \op B |\omega_0\rangle] = m[(\langle \omega_0| \op B^* \op U(t) \op A^* |\omega_0\rangle)^*] = (\langle \omega_0| \op B^* \op P_{\text{inv}} \op A^* |\omega_0\rangle)^* \\ = \langle \omega_0| \op A \op P_{\text{inv}} \op B |\omega_0\rangle
    \ee
    where the first equality is trivial and in the second equality we used \cref{eq:mean-ergodic-def} by taking $|\phi_1\rangle = \op B|\omega_0\rangle$ and $|\phi_2\rangle = \op A^*|\omega_0\rangle$. The third equality simply used the fact that being a projector $\op P_{\text{inv}}^* = \op P_{\text{inv}}$.
\end{remark}


\bibliographystyle{JHEP}
\bibliography{ref}      
\end{document}